# DynIMS: A Dynamic Memory Controller for In-memory Storage on HPC Systems

Pengfei Xuan, Feng Luo, Rong Ge, Pradip K Srimani
School of Computing
Clemson University
Clemson, SC, 29634 USA
{pxuan, luofeng, rge, psriman}@g.clemson.edu

*Abstract*— In order to boost the performance of data-intensive computing on HPC systems, in-memory computing frameworks, such as Apache Spark and Flink, use local DRAM for data storage. Optimizing the memory allocation to data storage is critical to delivering performance to traditional HPC compute jobs and throughput to data-intensive applications sharing the HPC resources. Current practices that statically configure in-memory storage may leave inadequate space for compute jobs or lose the opportunity to utilize more available space for data-intensive applications. In this paper, we explore techniques to dynamically adjust in-memory storage and make the right amount of space for compute jobs. We have developed a dynamic memory controller, DynIMS, which infers memory demands of compute tasks online and employs a feedback-based control model to adapt the capacity of in-memory storage. We test DynIMS using mixed HPCC and Spark workloads on a HPC cluster. Experimental results show that DynIMS can achieve up to 5X performance improvement compared to systems with static memory allocations.

## I. INTRODUCTION

Data-intensive computing workloads emerging on traditional HPC clusters [1] pose a great challenge on system memory management. Such workloads process large volumes of data often in terabytes or petabytes that are stored in persistent storage devices. In order to accelerate the speed of data access and processing, in-memory computing frameworks, such as Apache Spark and Flink, use local DRAM for storing data [2], [3]. The in-memory storage competes against the traditional HPC jobs on the limited physical DRAM capacity available on the system. Memory is a well-recognized key resource and is critical to HPC application performance. Limited capacity or poorly used memory system can lead to severely degraded performance for HPC applications. This memory conflict becomes even more problematic when mission-critical jobs simultaenously run on the HPC cluster.

Current practices that statically configure in-memory storage can't effectively address the memory conflicts between data-intensive workloads and traditional HPC workloads. Data-intensive computing frameworks, such as Spark, provide interface for users to specify a DRAM space as the in-memory storage. Though Spark can adjust the space partition between Spark workload execution and storage memory region within the framework [4], [5], it can't manage the space outside of the framework. In production environment, it is often hard or even impossible to determine a suitable configuration for memory allocation between compute jobs and data-intensive workloads. A small in-memory space may not be optimal for data-intensive workloads, while a large in-memory space may leave inadequate space to compute jobs and severely hurt their performance. Furthermore, memory demand varies significantly between compute jobs and during the execution of a single job.

A promising approach is to dynamically adjust the memory distribution between in-memory storage and execution runtime of traditional HPC job execution according to the latter's demand. This approach gives the priority to HPC compute workloads and meets their memory demands first. It can then opportunisitically allocate the rest available space to the data-intensive frameworks. In our prior work [6], we integrted a distributed in-memory storage system [7], [8] with a parallel file system to improve the performance of data-intensive jobs. However, the size of DRAM allocated for the in-memory storage system is fixed and the deployment of separated in-memory store reduces the memory size for compute jobs and hurts the performance of the overall job mix on the system. Built on our prior work, we explore to dynamically adjust in-memory data storage space at runtime to maintain HPC compute applications' performance and accelerate data-intensive workload execution.

In this work, we present *DynIMS*, a new dynamic memory controller to manage the capacity of in-memory storage system on HPC clusters. DynIMS can improve HPC system throughput when there are mixed compute- and data-intensive workloads. Specifically, our contributions are as follows:

1) We empirically investigate the impact of memory pressure on HPC workload performance.
2) We design a self-adaptive memory controller model, in which we use feedback control for dynamic capacity eviction and allocation of in-memory storage system.
3) We implement a prototype of DynIMS to control Alluxio [7], [8] in-memory storage system that is deployed on compute nodes of HPC cluster.

We evaluate DynIMS using mixed HPC cluster and Spark workloads and show up to 5× performance improvement over static memory allocation.

## II. BACKGROUND

### A. Memory Usage Pattern of HPC Applications

To understand the peak memory usage pattern of HPC applications, we run HPCC benchmarks of HPC workloads such as HPL, DGEMM, STREAM, PTRANS, RandomAccess, FFT, and a set of tests to measure networking bandwidth and latency. The peak memory usage of HPCC benchmark is close to 75 GB (Fig. 1). Thus, if we deploy a big data framework with a static configuration on compute nodes with 120 GB available memory that also run the HPCC benchmark, we can only have 25 GB space for in-memory storage and other 20 GB memory space for big data runtime and application execution. As shown in Fig. 1, at least 40 GB memory is unused during most of HPCC benchmark running time. The static configuration of in-memory storage leads to low usage of memory most of time.

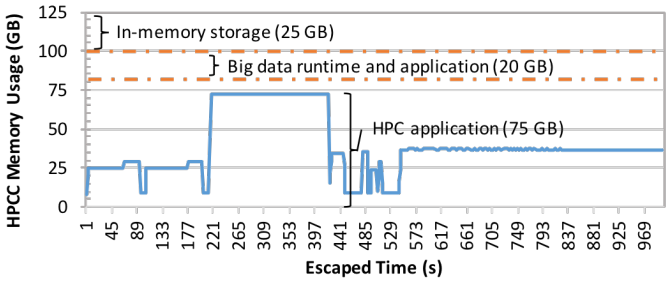

Fig. 1. Memory usage pattern on HPCC benchmark.

### B. Memory Pressure and HPC Application Performance

To understand the relationship between the memory pressure and the performance of HPC applications, we run the High Performance Linpack (HPL) benchmark on a single compute node with 24 CPU cores and 125 GB memory. The problem size of Linpack varies from 5 GB to 100 GB. When each Linpack instance is runing, we use other programs to control the overall system memory utilization and make it stay at certain levels. Once memory utilization reaches 100%, we can further request more memory to engage the swap space. In our experiments, we control the utilization of the swap space at 0.5% and 1% of the physical memory. Fig. 2 plots the measured performance of HPL benchmark at various system memory utilizations. HPL benchmark performance drops sharply as the system memory utilization is close to 100%. The performance behavior of HPL benchmark indicates that HPC applications are very sensitive to the memory pressure. Performance degradation or even application failure will happen if memory pressure is not released timely. Therefore, a sub-second or even millisecond-level response is required to avoid execution exception.

## III. SYSTEM DESIGN

### A. Architecture Overview

Our dynamic memory controller, DynIMS, implements a runtime monitoring scheme and consists of four major building blocks (Fig. 3):

- **Monitoring Agents:** they collect the memory usage statistics. We use *collectd* [9] as our monitoring agents, for which we configure memory and Kafka plugins to collect and forward the memory metrics. JSON format is used to code structural information.
- **Stream Processor:** it computes the optimized in-memory storage space for each node online and is

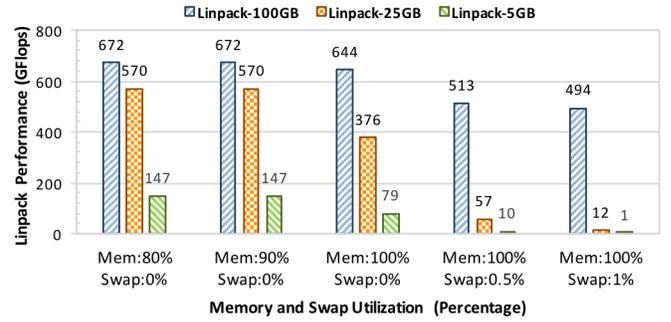

Fig. 2. Performance impact on system memory pressure.

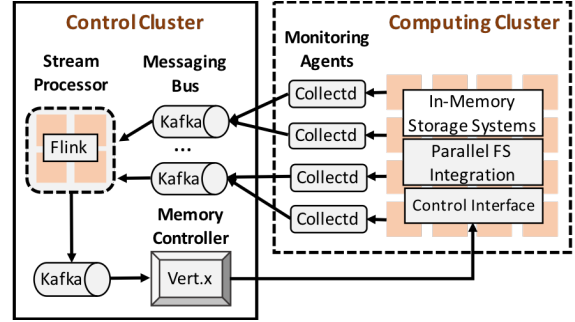

Fig. 3. DynIMS system architecture.

powered by Apache Flink [10]. The stream processor is implemented as a stream service and is scaled to whole control cluster. Stream processor includes a simple and flexible interface to programmatically interact with the aggregated memory metrics.
- **Memory Controller:** it determines and sends out the memory eviction and allocation instructions. Its implementation is based on Vert.x [11] framework. We also implement communication adapters and control interfaces between memory controller and in-memory storage.
- **Messaging Bus:** it transports the memory usage metrics and aggregated statistics. Apache Kafka [12], a distributed messaging system, is adopted to build messaging bus bridging above three modules.

The design of DynIMS emphasizes generality, modularity, and scalability. The DynIMS can provide an out-of-the-box solution to support majority in-memory storage systems. Memory controller is driven by a self-adaptive control model to dynamically regulate in-memory storage capacity. The input of DynIMS is a sequence of real-time memory usage metrics collected by each of monitoring agents, and the output of DynIMS is corresponding memory capacity adjustment instructions for in-memory storage on each compute node. Monitoring agents (collectd daemons) are distributed to each compute node for cluster-wide memory usage monitoring and forwarding. To enable the dynamic memory adjustment of in-memory storage systems on the runtime, a control interface is implemented based on file systems' APIs through RPC or REST interface. We have implemented two interfaces for Alluxio and HDFS.

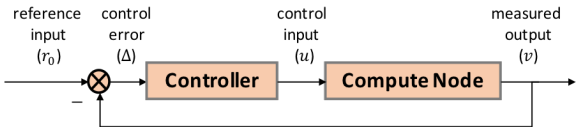

Fig. 4. Feedback-based control on memory adjustment.

DynIMS is carefully architected to reach a sub-second response for accommodating a burst of memory usage. To deliver a low-latency control cycle across the whole computing cluster, each of these four building blocks includes a scalable architecture to guarantee processing time. The performance of the messaging bus and stream processor relies on the underlying Kafka and Flink clusters. Both of those two frameworks are well proven for delivering throughput in the order of one and even tens millions of events per second with latency as low as few tens of milliseconds [13], [14]. In addition, the memory controller uses the event-driven and non-blocking architecture provided by Vert.x to handle high concurrency adjustment signal with a low cost, and can be scaled to multiple machines using Hazelcast or JGroups based clustering techniques. As a result, DynIMS can run very efficiently with a low overhead on memory monitoring and model computation. In our tests, the average computation cost on the aggregated metrics received from 4 compute nodes is below 10% utilization of a single CPU core.

### B. Memory Control Model

As discussed in Section (II.B), the utilization of memory has a close correlation with the application performance as well as system stability, and it is desirable to maintain the memory pressure below a critical threshold. We apply a feedback-based control model to adjust in-memory storage (Fig. 4). We continuously monitor the memory usage of each compute node. The usage information drives a controller to compute the next optimized size for in-memory storage. The controller signals the compute node to adjust in-memory storage.

Let $u_i$ and $v_i$ be the capacity of in-memory storage and system memory usage of a compute node with total memory size $M$ during the $i$th control interval. In addition, $r_i = v_i/M$ denotes the memory utilization ratio of a computer node in the same interval. The memory controller computes the suitable in-memory storage capacity for the next $(i+1)^{\text{st}}$ interval using the following equation:

$$u_{i+1} = u_i - \lambda v_i \frac{r_i - r_0}{r_0} \quad (1)$$

where $r_0$ is the threshold of memory utilization ratio on the compute node, and the $\lambda$ is a parameter that controls the aggressiveness of the tuning on in-memory storage. The value of $\lambda$ is related to speed at which the applications consume memory as well as the adjust interval, $T$, of controller.

It is important to choose the right $\lambda$ to make the feedback-based control system stable and the memory capacity of in-memory storage reduced to threshold value $r_0$ as quickly as possible. We have empirically evaluated the stability of DynIMS under a range of $\lambda$ ($0 < \lambda \leq 2$) against a fixed memory utilization threshold ($r_0 = 95\%$), and found that $\lambda = 0.5$ can delivery a good balance between stability and responsiveness on our testing workload.

TABLE I. PARAMETER VALUES OF THE MEMORY CONTROLLER.

| $M$ | $r_0$ | $\lambda$ | $U_{min}$ | $U_{max}$ | $T$ |
|---|---|---|---|---|---|
| 125 GB | 0.95 | 0.5 | 0 GB | 60 GB | 100 ms |

TABLE II. HARDWARE CONFIGURATIONS OF SELECTED NODES ON PALMETTO CLUSTER.

| CPU | Intel Xeon E5-2680 v3 24×2.50 GHz |
|---|---|
| HDD | 1 TB 7200RPM SATA |
| RAID | 12 TB LSI Logic MegaRAID SAS |
| RAM | 125 GB DDR3-1600 |
| Network | Intel 10 Gigabit Ethernet |
| Switch | Brocade MLXe-32 with 6.4 Tbps backplane |

To ensure that the adjusted in-memory storage size is within a machine-specified range, we define $U_{min} \leq u_{i+1} \leq U_{max}$, where $U_{min} = 0$ and $U_{max} = \alpha M$, $0 < \alpha < 1$; $\alpha$ is a machine-specific parameter. In addition, the control interval $T$ is also a very important parameter that directly affects system performance and stability. To maintain a high sensitivity to the memory pressure, we set the control interval as small as possible while keeping the monitoring and adjusting overheads within a reasonable range.

## IV. RESULTS

### A. Experiment Setup

In this section, we evaluate DynIMS on the Palmetto HPC cluster at Clemson University. Table I lists the parameters of DynIMS used in our experiments. We test DynIMS while running the Spark applications and HPCC benchmark simultaneously. We use Alluxio as our in-memory storage, which uses the RAMdisk as storage media. We apply LFU eviction policy on Alluxio backed by the OrangeFS parallel file system to form a two-level storage system [6].

We select nodes with the same hardware configuration (Table II) for our experiments to get a consistent test environment. Each compute node has a single 1 TB SATA hard disk and 60 GB RAMdisk; maximum capacity of Alluxio cannot be more than 60 GB in compute nodes. Each data node is equipped with 12 TB disk array backed by 80 GB OS buffer cache. All nodes are connected through 10 Gigabit Ethernet network. Although we cannot control the bandwidth of switch backplane, the backplane bandwidth is several orders of magnitude higher than the aggregated network throughput and thus is not the bottleneck resource in our experiments.

We run experiments on five compute nodes and a 2-node OrangeFS storage cluster. We select one compute node as the head node to host system services, including Spark Master, Alluxio Master, Flink Master and Worker, Kafka broker, Zookeeper server, and DynIMS controller. We deploy Spark executors, MPI runners, Alluxio Workers, and collectd as Kafka producer on each compute node.

The peak execution memory required by HPCC workloads is about 75 GB on each compute node and each Spark executor requires at least 20 GB execution memory to avoid the extra overhead caused by frequent JVM garbage collection (GC). Therefore, after other 5 GB reserved space to prevent memory pressure, there are only 25 GB available memory left for data storage on each compute node during the peak memory execution time of HPCC. Thus, 25 GB is the memory capacity we can assign to Alluxio with static configuration.

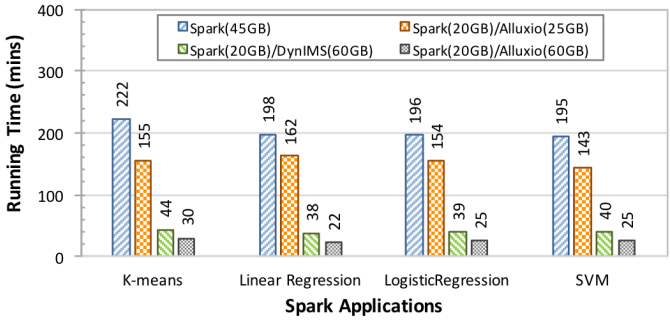

Fig. 5. Running time of different machine learning applications (320 GB datasets) using different memory configurations.

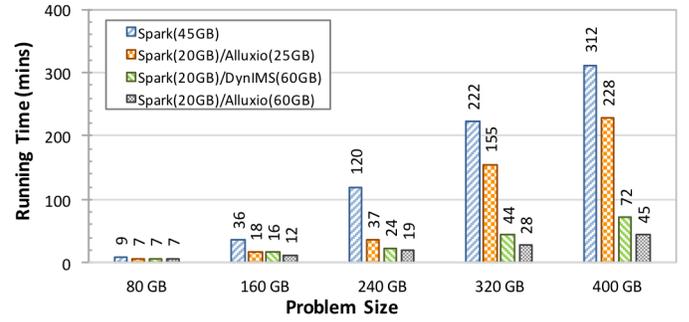

Fig. 6. K-means application with different problem sizes using different memory configurations.

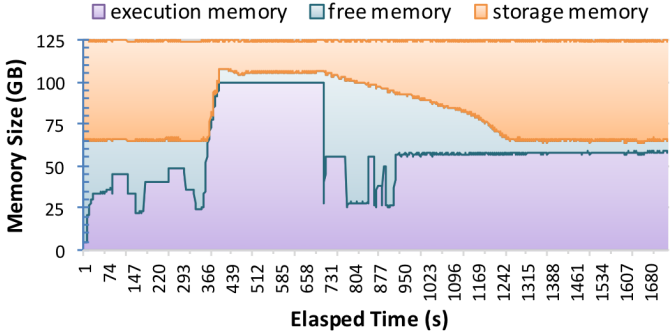

Fig. 7. The system memory statistics of K-means (320 GB dataset) and HPCC workloads during the peak memory demand.

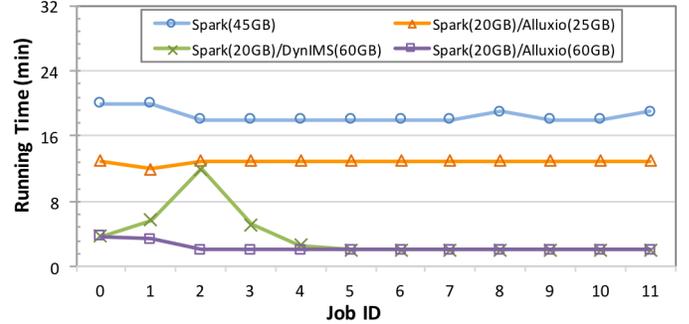

Fig. 8. Iteration time on K-means workload (320 GB datasets) using different memory configurations.

We run each experiment with four memory configurations:

**Configuration 1, Spark(45GB).** We assign 45 GB total memory for both execution and storage of Spark on each compute node; the data is read from OrangeFS through Alluxio without caching. This configuration is static and reserves about 25 GB for RDD caching that is immune to being evicted by execution (spark.memory.storageFraction = 0.56).

**Configuration 2, Spark(20GB)/Alluxio(25GB).** We assign 20 GB execution memory for Spark and offload the rest of 25 GB to Alluxio, and the data is read from OrangeFS and cached in Alluxio. This configuration is also static.

**Configuration 3, Spark(20GB)/DynIMS(60GB).** We assign all 60 GB RAMdisk to Alluxio initially. At runtime, we run DynIMS to adjust the capacity of Alluxio dynamically.

**Configuration 4, Spark(20GB)/Alluxio(60GB).** We assign all 60 GB RAMdisk to Alluxio. Different from Configuration 3, here we do not run HPCC benchmark and thus no memory burst occurs. This configuration delivers the upper bound of Spark application performance and serves as the reference for evaluating the efficiency and effectiveness of DynIMS.

### B. The Performance of DynIMS on Different Spark Applications

To evaluate the performance of DynIMS, we ran four different Spark applications: K-means, logistic regression, linear regression, and support vector machine (SVM).

We run each experiment with 10 iterations against 320 GB datasets using Hadoop SequenceFile format. For each experiment, we start the HPCC and Spark workloads together once the input datasets have been generated. Fig. 5 shows the experimental results. With dynamic memory adjustment using DynIMS, the Spark applications run 5.1× and 3.8× faster than with those of two static configurations. The Spark applications with DynIMS have comparable performance with their performance reference upper bound. The speedups come from a better hit-rate on input datasets. DynIMS leads up to 75% in-memory hit ratio on compute nodes in most of execution periods through dynamic memory adjustment. Moreover, the high in-memory hit ratio can further increase the efficiency of OS buffer cache located in data nodes by migrating hot datasets from data nodes to compute nodes and thus can reduce on-disk access overhead occurring in data nodes. As a comparison, statically configured Alluxio only can reach at most 31% in-memory hit ratio on compute nodes and has to read at least 69% (220 GB) of dataset from remote data nodes. Because the data nodes only have 160 GB aggregated memory space and the remote data cannot fit into the OS buffer cache of data nodes. Thus, Spark workloads experience a significant I/O degradation [15], [16]. Lastly, running time with caching a portion of input datasets (100 GB) in Spark RDD is 1.3× slower than those with keeping it in Alluxio. This is because the size of deserialized SequenceFile is often larger than the size of its original data and needs more caching space; reducing the amount of data that can be cached in Spark RDD leads to a poor cache hit rate.

### C. Impact of Insufficient Storage Memory

If the capacity of in-memory storage on compute node is not large enough to hold all data, part of the data needs be stored in remote OS buffer cache or disk. In our next experiment, we scale the input data size from 80 GB to 400 GB for K-means application, and run it with four memory configurations. As shown in Fig. 6, with DynIMS, the K-means running time increases much slower than those with static configurations do. The K-means performance with static configurations using OrangeFS and Alluxio starts to experience a significant degradation when the problem sizes reach 160 GB and 240 GB respectively. Therefore, DynIMS is not only able to improve in-memory hit ratio, but also increases compute efficiency and scalability when the problem size scales up.

### D. Stability and Responsiveness of DynIMS Control Model

As shown in Fig. 1, the HPCC workloads exhibit a dynamic demand on memory resource within a small portion of burst area. We expect that DynIMS can detect and adapt such memory burst through its feedback-based control model in real time. The memory control model should be properly designed and its parameters should be correctly selected; otherwise, DynIMS may become too aggressive or overly sensitive to small noises in measurements or transients in the workloads, resulting in large oscillations in the capacity of in-memory storage.

To understand the stability and responsiveness of the control model in DynIMS, we have examined the system memory statistics of a mixed HPCC and K-means workload. Fig. 7 shows the statistics of execution memory, free memory and storage memory of compute nodes with DynIMS. Alluxio starts with a capacity of 60 GB, and then adaptively shrinks its capacity to maintain the memory usage below the predefined threshold (95%) when there is a memory burst in HPCC workloads. After the memory burst disappears, Alluxio recovers its capacity back to its initial size gradually. Lower variance of in-memory storage capacity indicates that the proposed control model has a good stability. Meanwhile, the closely correlated sizes between execution memory and storage memory show the evidence of fast response of DynIMS with the selected control parameters.

Fig. 8 shows the running time of different iterations of K-means with four different memory configurations. During the memory burst time, the running times of K-means iterations (iteration 1, 2, 3) using DynIMS increase to those of K-means iterations using static configured Alluxio (25 GB) gradually. After the memory burst disappears, the running time of K-means iterations using DynIMS recovers back to its upper bound. This demonstrates that DynIMS is able to maximize the system throughput after memory pressure is released.

## V. RELATED WORK

Recently, in-memory computing is becoming a key approach to reduce the overhead of on-disk access cost on data-intensive workloads. The National Energy Research Scientific Computing Facility (NERSC) did a comprehensive evaluation as well as enhancement to scale Spark [17] on traditional HPC systems [18]–[22]. The early version of Spark manages the memory space statically with an isolated execution and storage memory. Since Spark v1.6.0, an unified memory management [4] is introduced to eliminate the boundary between execution and storage. Xu et al. [5] enhanced this concept by adding DAG dependency-based eviction on RDD cache. However, the total amount of system memory used by Spark still must be statically determined when configuring the cluster, and cannot change during runtime. How to release system memory dynamically from Spark runtime remains an open question. DynIMS targets to build a more general framework to support both HPC and big data workloads and provides an out-of-the-box solution to execute a mixed workload on HPC systems.

In addition, there are a few other research projects directly related to in-memory storage systems. Pu et al. [23] studied and analyzed the strategies of fair allocation of multi-user shared memory systems. Uta et al. [24] demonstrated the performance improvement by dynamically scaling out and scaling in the cluster size of MemFS based-on application demand on memory resource. Jeong et al. [25] proposed a set of system primitives (APIs) to enable dynamical adjustment on allocated memory resource.

Caching optimizations for parallel I/O systems are not new and have been widely explored. Panache [26] added a scalable caching layer atop of GPFS, which can persistently and consistently cache data and metadata from remote storage cluster. Modern HPC file systems often use dedicated I/O nodes for integrated data buffering and I/O forwarding [27]. These techniques are orthogonal and complementary to DynIMS, which could utilize them for its private, auxiliary, or primary metadata servers.

Finally, Chen et al. [28] proposed an algorithm-level feedback-controlled adaptive (AFA) to improve flexibility and efficiency of data prefetching instead of data caching. AFA can dynamically determine an appropriate prefetching algorithms at runtime for different access patterns using data-access history cache (DAHC) [29], which is orthogonal and complementary to our work. DynIMS could utilize this strategy to select the optimized eviction algorithm adaptively.

## VI. CONCLUSSION

In this paper, we design, implement and evaluate a dynamic memory controller, DynIMS, for in-memory storage system to accelerate a mixed HPC and Spark workload on HPC systems. DynIMS detects memory contention between task execution and data storage in real time, and adaptively determines the optimized in-memory storage capacity with its feedback system, and enables a fine-grained control on memory allocation and eviction. This can improve the performance of Spark over HPC systems. Resulting from either a too small or a too large storage memory, the original static configured Spark can lead to a low resource sharing, or deprive other execution tasks from obtaining sufficient memory to compute efficiently. Performance evaluation of our DynIMS shows up to 5× improvement on mixed HPCC and Spark workloads across a range of problem sizes compared with the static configurations. In future, we plan to evaluate the performance of DynIMS using more big data analytics frameworks, such as, Apache Flink, Tez, Hadoop MapReduce, and Apache Apex. In addition, a more sophisticated cache management is under our consideration. We plan to make DynIMS orchestrate second-level cache hosted on data nodes through write hints.

ACKNOWLEDGMENT

We are grateful for the support from the OrangeFS community led by Dr. Walter B. Ligon at Clemson University. CloudLab and the Clemson Computing and Information Technology (CCIT) support HPC resources used in this research.